\documentclass[preprint,12pt,3p]{elsarticle}
\usepackage{epsfig,amssymb,amsfonts,amsmath,mathtools,bm,color}
\biboptions{sort&compress}

\usepackage{lineno,hyperref,slashed}
\modulolinenumbers[5]

\usepackage{graphicx}
\usepackage{orcidlink}

\usepackage{bm}
\usepackage{braket}
\usepackage{multirow}
\usepackage{dsfont}
\usepackage{mathtools}
\usepackage{xcolor}
\usepackage[normalem]{ulem}

\renewcommand{\vec}[1]{\mbox{\boldmath$#1$\unboldmath}}
\newcommand{\Tch}{T_{\rm ch}}

\newcommand{\be}{\begin{equation}}
\newcommand{\ee}{\end{equation}}
\newcommand{\bea}{\begin{eqnarray}}
\newcommand{\eea}{\end{eqnarray}}
\newcommand{\beas}{\begin{eqnarray*}}
\newcommand{\eeas}{\end{eqnarray*}}

\newcommand{\vex}{{\bm x}}
\newcommand{\vey}{{\bm y}}

\def\vec#1{\boldsymbol{#1}}

\begin{document}

\begin{frontmatter}

\title{A large size of the pion-like excitations in the stringy fluid above $T_{ch}$ is model independent and required by current algebra}

\author{L. Ya. Glozman}

\address{Institute of Physics, University of Graz, A-8010 Graz, Austria,
e-mail: leonid.glozman@uni-graz.at}

\begin{abstract}
Multiple lattice evidences support the existence of a
confining but chirally symmetric stringy fluid regime of QCD above the
chiral symmetry restoration temperature at $T_{ch} \simeq 155$ MeV.
This regime is characterized by an approximate chiral spin
symmetry and its extensions  which means that the propagating
excitations represent the chirally symmetric quarks connected into color singlets by the chromoelectric string. Clear $\pi,\pi'$ peaks above $T_{ch}$ were
extracted on the lattice from the spatial and temporal correlators, which become broader with temperature and disappear roughly at $3T_{ch}$. The meson-like excitations above $T_{ch}$ were studied within the manifestly confining and
chirally symmetric model. It has been demonstrated that the chiral symmetry restoration in the confining regime happens because of Pauli blocking of the
levels, required for the existence of the quark condensate, by the thermal
quark excitation. The same Pauli blocking leads to a huge swelling of the
low-spin mesons above $T_{ch}$ which become infinitely large in the chiral limit.
This property should be crucial for the explanation of the high collectivity and a very small mean-free path of the constituents above $T_{ch}$ seen experimentally. Here we demonstrate that the swelling of pions above $T_{ch}$
is a model-independent effect required by current algebra.
\end{abstract}


\end{frontmatter}


\section{Introduction}

Lattice studies indicated the emergence of approximate chiral spin $SU(2)_{CS}$ and $SU(4)$ symmetries in hot QCD matter above $\Tch$ thus suggesting that the matter should still be in the confining regime \cite{Rohrhofer:2017grg,Rohrhofer:2019qwq,Rohrhofer:2019qal,Chiu:2023hnm}.
This regime of QCD was called a stringy fluid, where the propagating
degrees of freedom are chirally symmetric quarks connected into color-singlets
by the electric string.
Also, there are additional not related to symmetry  evidences supporting the existence of this intermediate stringy fluid regime between the hadron gas and the quark-gluon
plasma \cite{Glozman:2022lda,Lowdon:2022xcl,Cohen:2023hbq,Cohen:2024ffx,Glozman:2025rhe,Mickley:2024vkm,Fujimoto:2025sxx},
for complementary reviews see \cite{G1,G2}.

The origin of the chiral symmetry restoration in the confining regime
as well the properties of the meson-like excitations have been studied \cite{Glozman:2024xll,Glozman:2024dzz}
within the manifestly confining and chirally symmetric model \cite{Amer:1983qa, LeYaouanc:1984ntu, Adler:1984ri,Bicudo:1989sh,Bicudo:2002eu,Llanes-Estrada:1999nat,Alkofer:2005ug, Wagenbrunn:2007ie,Quandt:2018bbu}. 
The nature of the chiral restoration in the confining regime is Pauli blocking
of the quark levels, required for the existence of the quark condensate,
by the thermal quark excitations. The same Pauli blocking leads to a huge
swelling of the low-spin mesons above $T_{ch}$ as compared to their size
in hadron gas. In the chiral limit these meson-like excitations become infinitely large in the confining regime. At the realistic quark masses the increase
of the low-spin meson sizes as compared to their size below $T_{ch}$ is of the 
order 5. This huge swelling of the meson-like excitations in the stringy fluid
should be crucially important for the properties of the medium: the medium
gets highly collective with a very small mean free path of its constituents.
The latter properties were observed experimentally to be crucial for the
hot matter at RHIC and LHC temperatures \cite{Heinz:2013th}.

Given these results the microscopic structure of the stringy fluid
could be shortly outlined as follows \cite{Glozman:2026glk}:

$\bullet$ It is a densely packed system of the overlapping color-singlet
large meson-like systems.
 
$\bullet$ It is a highly collective medium with a very small mean-free
path of the color-singlet constituents.

$\bullet$ There are no deconfined gluons.

$\bullet$ The propagating in time degrees of freedom are only color-singlets,
in which the massless quarks and antiquarks are connected by the chromoelectric string.

$\bullet$  Quark interchanges between the overlapping color-singlet meson-like systems,
required by Pauli principle, make fluctuations of conserved charges looking
as if the quarks were free.

Given  importance of the swelling of the low-spin meson-like excitations
in the stringy fluid as compared to the hadron gas, it is demanding to establish
whether it is a property of a model, or it is something more general. In this letter we demonstrate that such a large swelling is required by the current algebra and by the Gell-Mann-Oakes-Renner relation \cite{Gell-Mann:1968hlm} that can be applied
in the chiral symmetry broken phase at $T \leq T_{ch}$, i.e., in the hadron gas phase
with spontaneously broken chiral symmetry. 

The findings of this paper support the existence of the intermediate
stringy fluid regime \cite{Rohrhofer:2017grg,Rohrhofer:2019qwq,Rohrhofer:2019qal,Glozman:2022lda,Lowdon:2022xcl,Cohen:2023hbq,Cohen:2024ffx,Glozman:2025rhe,Mickley:2024vkm,Fujimoto:2025sxx}. However, they provide a qualitatively new microscopic insight into the structure of the stringy fluid which could not be addressed in previous studies.
\footnote{For example, the model study of Ref. \cite{Fujimoto:2025sxx} cannot address the effects  of chiral symmetry: it is not known how to implement chiral symmetry into the string description of hadrons, because there is no realistic string construction with massless quarks with spin at the ends of the string in 3+1 dimensions.}
We remind that in the chiral limit
the smooth chiral symmetry restoration crossover around $T_{ch} \sim 155$ MeV becomes a second order
phase transition at a temperature $T \sim 130$ MeV \cite{Karsch}. Consequently
in the chiral limit the stringy fluid is a phase of QCD just above the hadron
gas phase. At some high temperatures this phase should smoothly transform into
what is known as the quark-gluon plasma with deconfined quark and gluon
quasiparticles as degrees of freedom.

\section{Review of the meson lattice correlators above $T_{ch}$ and of
the confining and chirally symmetric model that predicts the swelling of mesons
above $T_{ch}$}

For pedagogical reasons and to establish a connection with the model-independent
statement about a large size of the pion-like excitations above the chiral restoration (that will
be discussed in the next section) we begin with the overview of the lattice meson correlators within $N_F=2$ QCD with the chirally symmetric Dirac
operator at physical quark masses \cite{Rohrhofer:2019qwq,Rohrhofer:2019qal}.

In Fig. 1  we show temporal correlators of the isovector
meson quark-antiquark bilinears with $J=0,1$: the scalar (S; $a_0$),
 pseudoscalar (PS; $\pi$) as well as $J=1$ correlators for two different types
 (with respect to chiral symmetry) of the $\rho$-correlators $\rho_{(0,1)+(1,0)}$,$\rho_{(1/2,1/2)_b}$ and $a_1$, $b_1$ correlators
 in QCD at $T=220$ MeV (right panel) and for noninteracting quarks (left panel).
\begin{figure}
\centering 
  \includegraphics[width=0.49\linewidth]{{{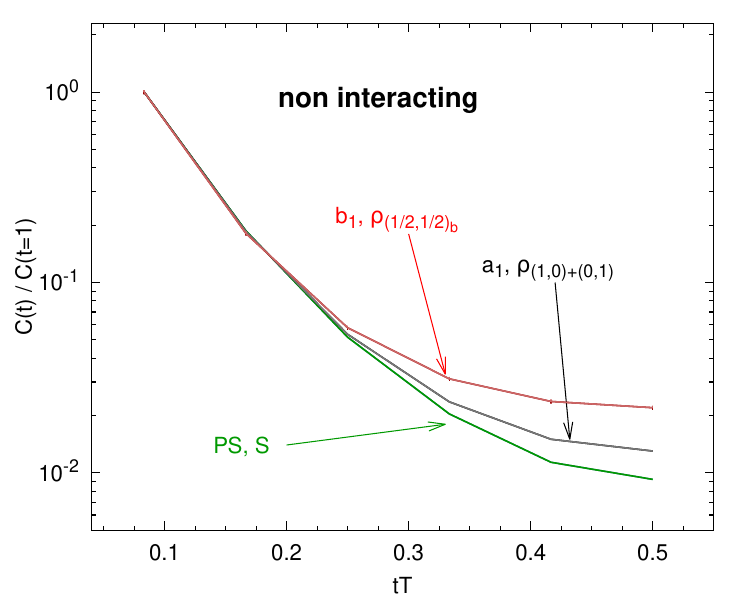}}}
  \includegraphics[width=0.49\linewidth]{{{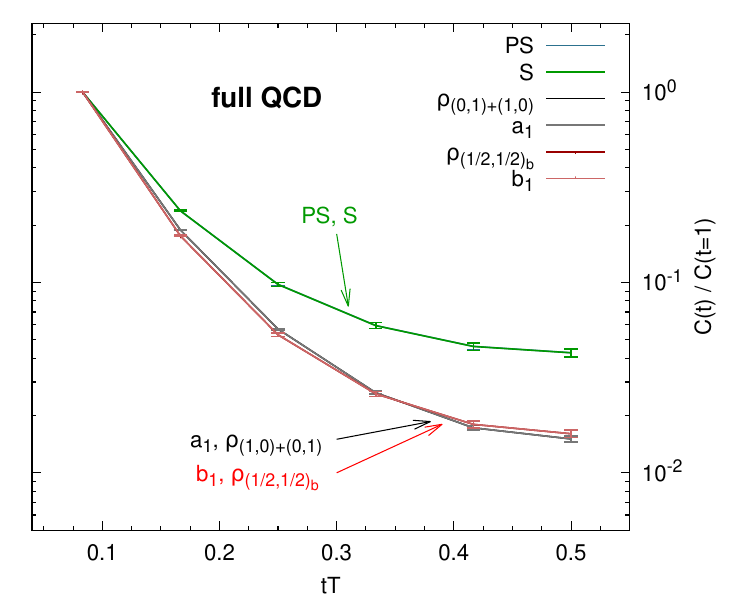}}}
    \label{temporal}
\caption{Temporal correlation functions for $12 \times 48^3$
lattices. The l.h.s. shows correlators calculated with free
noninteracting quarks with manifest $U(1)_A$  and $SU(2)_L \times SU(2)_R$
symmetries. The r.h.s. presents full QCD results at a temperature 220 MeV,
which shows multiplets of all  $U(1)_A$, $SU(2)_L \times SU(2)_R$, $SU(2)_{CS}$  and $SU(4)$ groups. From Ref. \cite{Rohrhofer:2019qal}.
}
\end{figure}
The correlators in the left panel represent a free quark loop of Fig. \ref{loo}.
 In this case the $U(1)_A$ and $SU(2)_R \times SU(2)_L$ are manifest.
The degeneracy of the $S$ and $PS$ correlators in QCD above $T_{ch}$ reflects
at least approximate restoration of the $U(1)_A$ symmetry. The $a_1 - \rho_{(0,1) + (1,0)}$ degeneracy is due to the restored $SU(2)_R \times SU(2)_L$ and the $b_1 - \rho_{(1/2,1/2)_b}$ degeneracy evidences the $U(1)_A$.
The approximate
degeneracy of all $J=1$ correlators indicates the emergence of approximate
$SU(2)_{CS}$ and $SU(4)$ symmetries, which are symmetries
of the confining electric interaction in QCD. Hence the mesonic correlators
in QCD can be schematically depicted through the diagram on the r.h.s.
of Fig.\ref{loo}. We see a qualitative difference between the QCD correlators
and free quark gas correlators.

\begin{figure}
\centering 
  \includegraphics[width=0.35\linewidth]{{{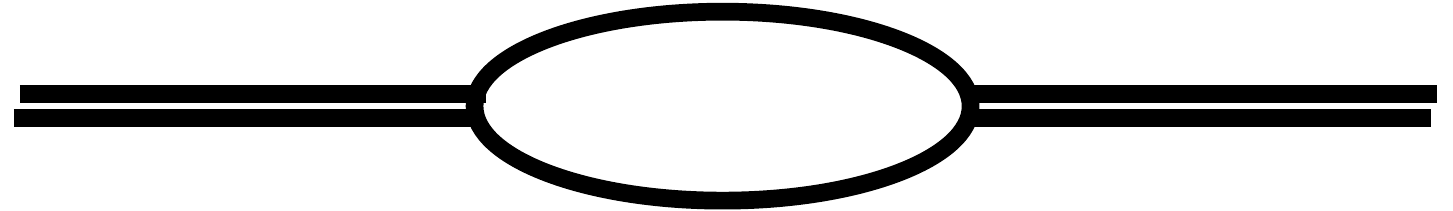}}}
  \includegraphics[width=0.35\linewidth]{{{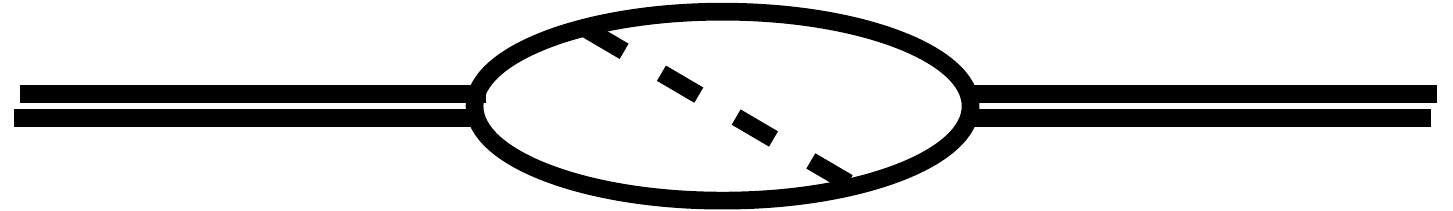}}}
  \caption{Left diagram - free quark loop correlator, right diagram -mesonic correlator with the dashed line representing the
  confining electric string.}
\label{loo}
\end{figure} 

\begin{figure}
  \centering
  \includegraphics[width=0.45\linewidth]{{{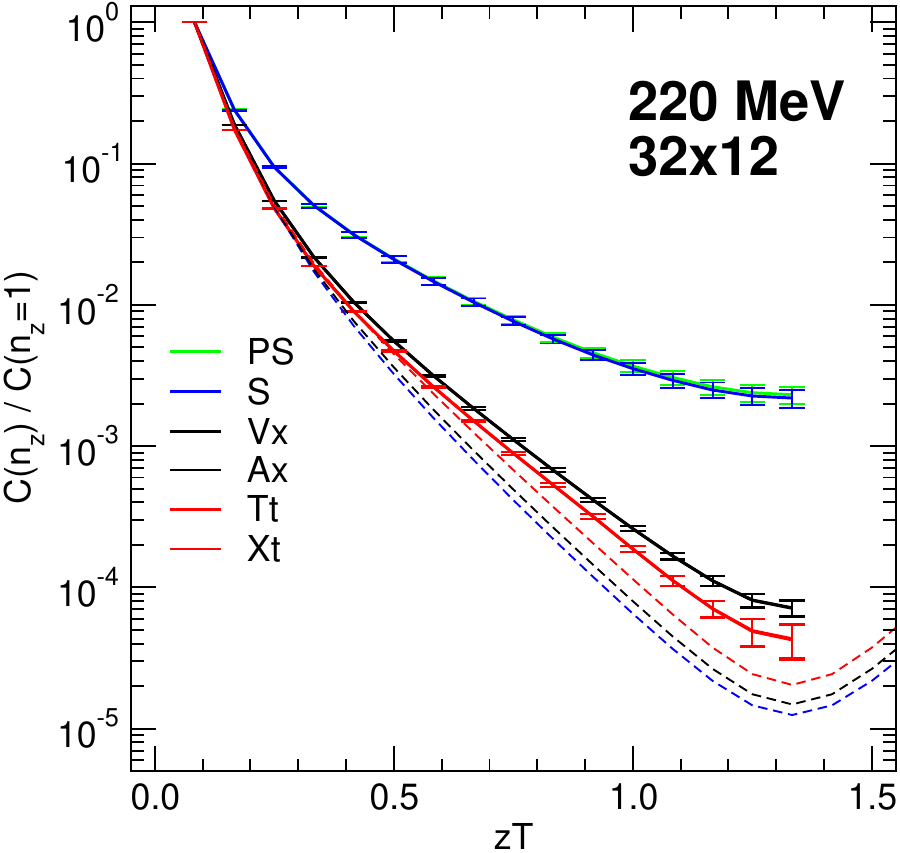}}} 
  \includegraphics[width=0.45\linewidth]{{{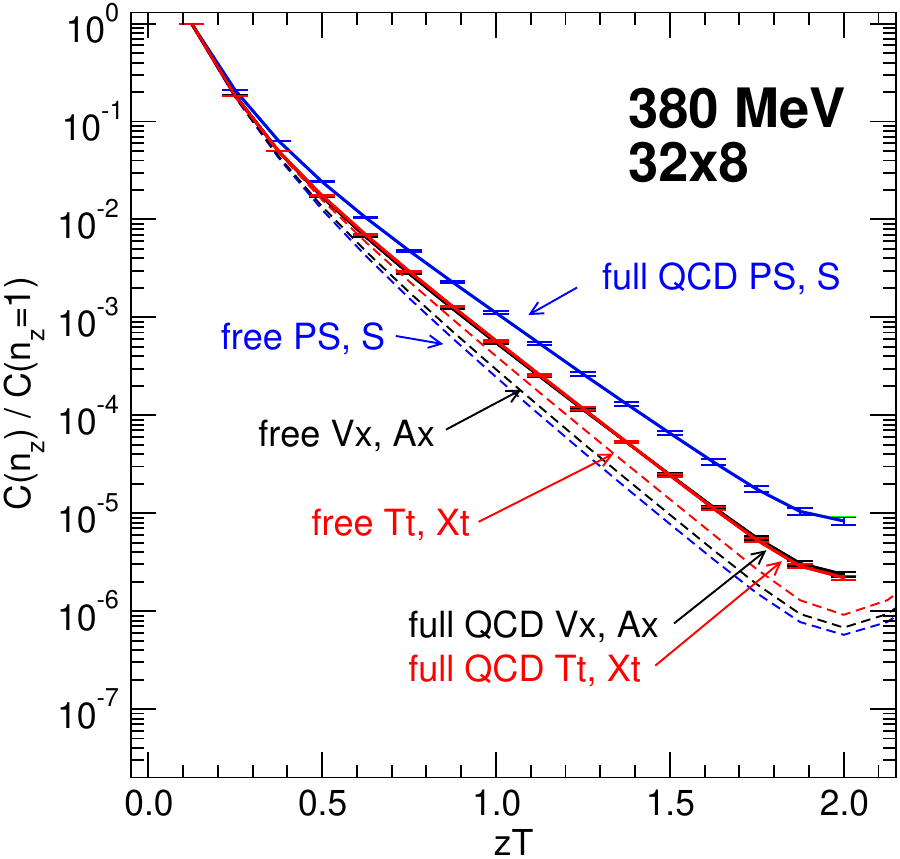}}} 
 \caption{Comparison of the full QCD spatial mesonic correlators (solid lines)
  with with the correlators obtained
  on the same lattice with free noninteracting quarks (dashed). From Ref. \cite{Rohrhofer:2019qwq}.}
\label{comparison}
\end{figure} 

In Fig. \ref{comparison} we show spatial isovector
$S$ and $PS$ correlators and $J=1$ isovector correlators. The $S -PS$ as well as the $T_t-X_t$ are connected by the $U(1)_A$ and the $V_x-A_x$
transform into each other by the $SU(2)_R \times SU(2)_L$. All four $J=1$ correlators
are connected by the $SU(2)_{CS}$ and $SU(4)$ symmetries.
Like in the case of the temporal correlators we again see the restored
$U(1)_A$ and  $SU(2)_L \times SU(2)_R$ symmetries and emerged approximate 
$SU(2)_{CS}$ and $SU(4)$ symmetries. The latter symmetries indicate
a confining regime and suggest the physics depicted on the r.h.s. of Fig.  \ref{loo}.
The  QCD correlators (solid lines) and
the  free quark loop correlators (dashed), obtained on the same lattice,
can be compared. Again we see a qualitative difference between the QCD correlators
and the free quark loop correlators.
 For the purposes of the present paper
please pay attention to a dramatic difference of the pion correlators ($PS$)
in QCD and the free quark gas. The spatial and temporal $PS$ correlators were
used to extract the pion spectral function above $T_{ch}$  \cite{Lowdon:2022xcl}
that shows very clear
$\pi$ and $\pi'$ peaks that become broader with temperature and disappear
roughly at $3 T_{ch}$.

\bigskip

Now we will overview a model that predicts a huge swelling
of the low-spin meson states above $T_{ch}$ as compared to
their vacuum size.
The model Hamiltonian is  the QCD Hamiltonian in the Coulomb
gauge where its gluonic part retains only  the instantaneous confining  term 
\be
\begin{split}
H=&\int d^3x\;\psi^\dagger(\vex,t)\left(-i\vec{\alpha}\cdot
{\bm\nabla}+\beta m\right)\psi(\vex,t)\\
& + \frac{1}{2} \int d^3x\; d^3y\;\rho^a(\vex) V_{conf}(|\vex-\vey|)\rho^a(\vey),
\label{GNJL}
\end{split}
\ee
which includes the interaction of two quark color charge densities,
$\rho^a=\psi^\dag\frac{\lambda^a}{2}\psi$, taken at the spatial points $\vex$ and $\vey$, via an instantaneous linear confining potential.
The quark kinetic part is chirally symmetric while the confining  part is invariant under larger symmetry groups: $SU(2)_{CS}$, $SU(2N_F)$, and $SU(2N_F) \times SU(2N_F)$ \cite{G1,G2}.

The chiral symmetry breaking at low temperatures is obtained via
solution of the Schwinger-Dyson
gap equation in the rainbow approximation. The chiral symmetry restoration phase
transition happens because of Pauli blocking of the levels, required for the existence
of a nonvanishing quark condensate, by the thermal excitations of quarks \cite{Glozman:2024xll}.
The quark-antiquark
excitations at different temperatures follow from the Bethe-Salpeter equation
in the ladder approximation  \cite{Glozman:2024dzz}. While the model was solved at $N_c=3$, the rainbow-ladder is strictly valid at large $N_c$.

The solution of the Bethe-Salpeter equation consists of the propagating forward
and backward in time quark-antiquark
"wave functions" $\psi_+(p)$ and $\psi_-(p)$. They are normalized according to
\be
\int\frac{p^2dp}{2\pi^2}\Bigl[\psi_+^2(p)-\psi_-^2(p)\Bigr]=2m_{\pi},
\label{norm2M}
\ee
 which follows from the condition that the charge of the state with $I=I_3=1$ equals to unity. Here $m_{\pi}$ is the energy of the pion-like state obtained
 from the solution of the Bethe-Salpeter equation.

\begin{figure*}[t!]
\centering
\includegraphics[width=0.5\textwidth]{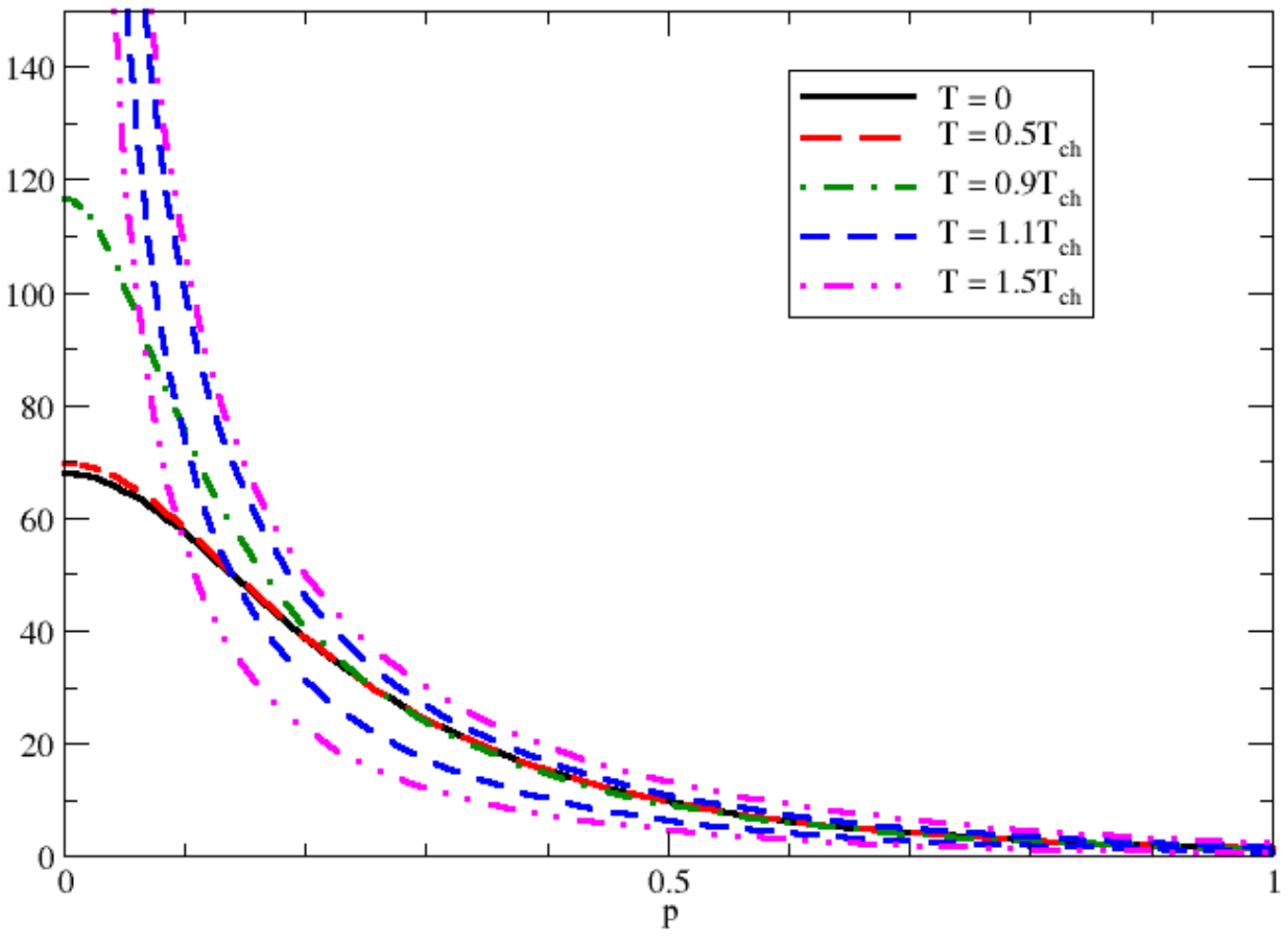}
\caption{The ground state ($n=0$) wave functions $\psi_\pm(p)$ for $J^{PC}=0^{-+}$   at different temperatures. For $J^{PC}=0^{-+}$ and $T<\Tch$, the corresponding pseudoscalar meson is a massless Goldstone boson with $\psi_+(p)=\psi_-(p)$. At $T>\Tch$  $\psi_+(p)>\psi_-(p)$, so for each temperature there are two curves. All dimensional quantities are given in the
appropriate units of $\sqrt \sigma$, where $\sigma$ is the Coulomb string tension. From Ref. \cite{Glozman:2024dzz}.}
\label{fig:wfs}
\end{figure*}

In Fig. \ref{fig:wfs} we demonstrate the "wave functions"  $\psi_{\pm}(p)$  for the ground color-singlet  quark-antiquark systems with pion quantum numbers $J^{PC}=0^{-+}$ 
at different temperatures in the chiral limit.
At $T=0$ and all temperatures $T < T_{ch}$ the pion "wave functions" are
localized in a small space volume. This is a result of a confining interaction
between quarks that acquire a dynamical mass  due to the spontaneous breaking
of chiral symmetry.  At the chiral restoration temperature
$T_{ch}$ the dynamical quark mass vanishes and the color-singlet quark-antiquark system
gets delocalized. This delocalization can be seen from the divergence
of 
the wave functions $\psi_+(p),\psi_-(p)$ at $p=0$ and the
system acquires infinite root-mean-square radius in the chiral limit.
At realistic quark masses the root-mean-square radius of the pion-like excitations obtained from the pion formfactor,  increases
by a factor $\sim 5$ as compared to the one below $T_{ch}$. Of course, this factor $\sim 5$ is specific to the model, where many possible QCD effects
are missing. In addition, it is difficult to estimate the $1/N_c$ corrections
within this model as it would require to go beyond the rainbow-ladder
approximation.  Notice that
the delocalized color-singlet quark-antiquark system  knows about confinement: there are no color-triplet free quark poles and the spectrum
of the color-singlet  quark-antiquark systems is discrete.

The huge swelling of the low-spin "mesons" and in particular of the pion above $T_{ch}$ has significant
phenomenological implications. The hot QCD matter above $T_{ch}$ is a dense medium of overlapping huge color-singlet quark-antiquark
systems ("strings"). Consequently the matter is highly collective and the mean free path of the  color-singlet constituents approaches zero.

\section{A model-independent derivation of a large swelling of the 
pion-like excitations above $T_{ch}$}

Here we argue that the results of the previous section about a huge
size of the pion-like excitation in the chirally symmetric confining phase, obtained within a model,
are actually model-independent. The argument is based on the well-known result
of current algebra for pion mass in the Nambu-Goldstone mode of chiral
symmetry, which is known as the Gell-Mann-Oakes-Renner relation (GMOR) \cite{Gell-Mann:1968hlm}. 

The correlation function of the isovector axial-vector current in the
Nambu-Goldstone mode of chiral symmetry  is dominated by the pion pole,
which is massless in the chiral limit:

\be
i \int d^4x e^{ipx} <0|T \{ A^\mu_i(x) A^\nu_k(0)\}|0> = \frac{F_{\pi}^2 \delta_{ik}}{m_{\pi}^2 - p^2} p^\mu p^\nu
 + ...
\ee
where $F_{\pi}$ is the weak pion decay
 $\pi^+ \rightarrow \mu^+ \nu_\mu$ constant that parameterizes the matrix
 element (the Bethe-Salpeter amplitude)
\be
 <0|\bar d(x) \gamma^\mu \gamma_5 u(x)|\pi(p)> = i \sqrt{2}p^\mu F_{\pi} e^{-ipx};~~~F_{\pi}\simeq 92~ MeV.
\label{fpi}
\ee 
Physically the weak pion decay constant measures the  amplitude
of the quark and antiquark inside the pion to be at the same spatial point (at the origin). This is because the weak decay vertex $\bar q q W^\pm$
is point-like.
 
 It is instructive to remind the reader some points of the derivation
 of the GMOR formula. In the Nambu-Goldstone mode the pion state can be created from
 the vacuum both by the axial-vector current (\ref{fpi}) and by the PS bilinear,
 $<0|\bar d(x) \gamma_5 u(x)|\pi(p)> \neq 0$. The partial conservation of the axial vector current connects the 4-divergence of the axial vector current with the PS bilinear times $(m_u + m_d)$. While the Bethe-Salpeter amplitude (\ref{fpi})
 is directly sensitive to the weak decay of the pion, the amplitude 
 $<0|\bar d(x) \gamma_5 u(x)|\pi(p)>$ is not.
  
The GMOR relation reads
\be
m_{\pi}^2 F_{\pi}^2 = - (m_u + m_d) <0|\bar q q|0> + O(m_q^2),
\label{GMOR}
\ee
where $m_{\pi}$ and $m_q, ~~~q = u,d$ are pion and quark masses,
$<0|\bar q q|0>$ is the quark condensate. 
 Both the quark condensate and the pion decay constant
are order parameters for the spontaneous breaking of chiral symmetry. 

In a dilute hadron gas below $T_{ch}$ the pion state can be defined as in vacuum.\footnote{At large $N_c$ mesons do not interact and the dilute meson gas approximation at finite $T < T_{ch}$ gets exact.}
Hence the GMOR formula can be derived as in vacuum. In addition one can
use the finite temperature definition of the  pion state \cite{Bros:2001zs,Lowdon:2022xcl} that satisfies the
finite temperature extension of the K\"allen-Lehman spectral representation.
At a finite temperature in the hadron gas the GMOR formula keeps its vacuum
form, as it follows from the chiral perturbation theory \cite{Gasser:1986vb,Toublan:1997rr}.

At the chiral symmetry restoration temperature both
the quark condensate and the weak pion decay constant  vanish.
Above the chiral restoration point, i.e., in the Wigner-Weyl mode of chiral symmetry, the GMOR formula cannot be derived and it does not constrain the pion-like excitation energy in the chirally symmetric phase: the energy of the pion-like
excitation, if it exists, apriori can take any value. 

For us it is important that the weak pion decay constant, as an order parameter of chiral symmetry breaking, vanishes
at $T_{ch}$. This fact is clearly seen from the degeneracy of the vector
and axial-vector correlators, discussed in sec. 2, because a nonvanishing $F_\pi$
would necessarily split the  vector and axial-vector correlators. 

This can happen in two cases: 

(i) The quark and antiquark in the chirally symmetric phase are not
correlated plane waves. This is typical for the quark-gluon plasma where
deconfined quarks and gluons satisfy perturbation theory.

(ii) The quark and antiquark form a correlated pion-like state (bound state or resonance) with a huge size so that the Bethe-Salpeter amplitude (\ref{fpi}) for the
quark and antiquark to be at the origin vanishes. This is the case discussed within the model in the previous section. This model satisfies the requirement
that $F_\pi=0$ at $T \geq T_{ch}$.

The possibility (i) can be ruled out by the lattice PS correlators above
$T_{ch}$, see section 2. The difference between the QCD pion correlators
and the free quark loop is so huge so that the QCD correlator cannot be described
by  the perturbation theory, which is generically applicable only when the
difference between the QCD correlator and the free quark loop is small.
In addition, the emerged chiral spin and $SU(4)$ symmetries are also incompatible
with the perturbation theory. 

This situation implies that only the case (ii) applies to QCD above $T_{ch}$
and below the deconfinement temperature $T_d$.
The PS correlator contains  above $T_{ch}$ some pion-like intermediate states. For a possible realization of these states see Ref. \cite{Lowdon:2022xcl}.
However, these states must have such structure so they do not 
see the axial-vector current at the origin.\footnote{Schematically their "wave function" vanishes at the origin while the pion "wave function" at $ T < T_{ch}$
looks like a Gaussian (since L=0) with a finite amplitude at the origin.}
This implies that these states must have a very big size.

The suggested argument means that the appearance of the huge pion-like
excitations above $T_{ch}$ is a model-independent statement based on current algebra and lattice data. However, the present argument does not allow to
quantify how big is the swelling of the pion-like excitation above $T_{ch}$.
One needs dedicated lattice studies to determine the size of these
excitations about  and above $T_{ch}$. These studies should measure the formfactor of the pion-like excitation. However, an easier estimate of
the size of this object could be obtained from the variational calculation of this state with a few different gauge-invariant spatial smearings of the local
PS bilinear.

\section{Conclusions}

In this paper we have presented a model-independent statement that
the pion-like excitations above the chiral symmetry restoration transition
are the correlated states (bound states or resonances) with the essentially
larger size than in the hadron gas phase. The argument is based on the validity
of the current algebra and of the Gell-Mann--Oakes-- Renner relation in the
chiral symmetry broken phase below the chiral restoration transition and
on QCD correlation functions obtained above $T_{ch}$.
While the statement does show that the size of the pion-like excitations
above $T_{ch}$ is essentially (much) larger than the size of pions in hadron
gas, it cannot quantify precisely how much. The size of the meson-like excitations
is crucially important to explain the properties of the stringy fluid regime
(phase) above the chiral restoration transition. One needs  dedicated
lattice studies to clarify this issue.

\section{Acknowledgments}

The author thanks Tom Cohen, Aleksey Nefediev and Owe Philipsen for a careful reading of the 
manuscript and consequent discussions.
The  research is supported through the grant PAT3259224 of the Austrian Science Fund (FWF).

\end{document}